# Control of the chirality and polarity of magnetic vortices in triangular nanodots

M. Jaafar[1,], R. Yanes[1], D. Perez de Lara[2], O. Chubykalo-Fesenko[1], A. Asenjo[1*], E.M. Gonzalez[2], J.V. Anguita[3], M. Vazquez[1] and J.L. Vicent[2]

[1]Instituto de Ciencia de Materiales de Madrid, CSIC, Sor Juana Inés de la Cruz, 3, Cantoblanco, 28049, Madrid, Spain

[2] Departamento Física de Materiales, Facultad CC. Físicas, Universidad Complutense, 28040 Madrid, Spain

[3]Instituto de Microelectrónica de Madrid, CNM-CSIC, Isaac Newton 8, Tres Cantos, 28760 Madrid, Spain

AUTHOR EMAIL ADDRESS (aasenjo@icmm.csic.es)

ABSTRACT

Magnetic vortex dynamics in lithographically prepared nanodots is currently a subject of intensive research, particularly after recent demonstration that the vortex polarity can be controlled by in-plane magnetic field. This has stimulated the proposals of non-volatile vortex magnetic random access memories. In this work, we demonstrate that triangular nanodots offer a real alternative where vortex chirality, in addition to polarity, can be controlled. In the static regime, we show that vortex chirality can be tailored by applying in-plane magnetic field, which is experimentally imaged by means of Variable-Field Magnetic Force Microscopy. In addition, the polarity can be also controlled by applying a suitable out-of-plane magnetic field component. The experiment and simulations show that to control the vortex polarity, the out-of-plane field component, in this particular case, should be higher than the in-plane nucleation field. Micromagnetic simulations in the dynamical regime show that the magnetic vortex polarity can be changed with short-duration magnetic field pulses, while longer pulses change the vortex chirality.

KEYWORDS (Word Style "BG_Keywords"). Magnetic Vortex, Micromagnetic simulations, Magnetic Force Microscopy, Magnetic nanostructures, Reversal Magnetization.

## I. INTRODUCTION

Lithographed magnetic nanostructures offer an abundant possibility for tailoring their properties through the suitable choice of their geometry that determines the strength of shape anisotropies with magnetostatic origin[1]. The nanostructuring techniques have opened a road for the discovery of new physical phenomena at the nanoscale with important consequences for nanotechnological applications. Magnetic nanostructures present as valuable candidates for the development of high density storage media, high speed magnetic random access memories, Nanoelectromechanical systems (NEMS) and magnetic sensors and logic devices[2]. Arrays of nanoelements exhibit different magnetic behaviour as a function of their size, aspect ratio and separations[3,4]. Particularly, the ground state magnetic configuration of a nanodot evolves from a single domain state to a vortex state by increasing its size[5,6]. Since its first observation by means of the X-ray magnetic circular dichroism technique[7,8], the experimental study of vortex dynamics has become possible and has immediately attracted a lot of attention with the aim to control this process.

The vortex ground state exists in a wide range of nanodot sizes from several nanometers to microns depending on the interplay between the magnetocrystalline anisotropy, exchange and magnetostatic energies[5,9]. It is characterized by the *polarity* -the up or down direction of the vortex core magnetization- and the *chirality* -clockwise or counter-clockwise magnetization rotation. The resulting four possible states are independent and, consequently, the vortex state could store the information of four magnetic bits. It is known that the energy barrier separating the two states with opposite polarities is a Bloch point with an excess of exchange energy[10] and consequently, the vortex is remarkably stable against thermal fluctuations. At the same time, it has been recently discovered that the vortex polarity could be switched with very small perpendicular field and current pulses[8,11,12] via the mechanism of the vortex-antivortex pair creation[13,14,15]. This has strongly stimulated the idea of the use of the vortex state in non-volatile high density magnetic storage media[11,12,16] and vortex magnetic random access memories (VRAM)[17]. All previously reported analysis on the switching of the polarity of the vortex core is devoted to the circular dots where the chirality is not visible by Magnetic Force Microscopy (MFM) and cannot be controlled. Alternatively, it has been reported that in circular dots with truncated edges and other certainly complex engineered defects[18-25] the nucleation point of the vortex could be controlled. This allows a possibility of additional control of vortex chirality. Here we show that in a simple triangular dot geometry both vortex polarity and chirality can be tailored by suitable application of magnetic fields. Notice that the use of this quite simple triangular geometry provides an additional advantage of breaking the circular symmetry so making the chirality and the polarity directions well distinguishable by MFM technique[26]. This also offers a possibility of signal codification.

In the present paper, we discuss several possibilities aiming to achieve a complete control of both vortex polarity and chirality in triangular dots. We illustrate the concept of the vortex states control by applying in-plane and out-of plane magnetic fields in a static regime. Complementary micromagnetic simulations reveal the mechanisms of chirality and polarity control by applying field pulses. Additionally, the micromagnetic simulations provide us the information on the minimum field strength and duration requirements.

## II. VORTEX CONFIGURATION CONTROL BY IN-PLANE FIELD

### A. Experimental results

In this work, triangular Ni nanostructures have been prepared by nanolithography techniques. Their topography was characterized by SEM (Scanning Electron Microscopy) and AFM (Atomic Force Microscope), while the magnetic behaviour was studied by VFMFM (Variable Field Magnetic Force Microscopy). Square arrangement of Ni(111) triangles have been fabricated by electron beam lithography and magnetron sputtering techniques [27]. The triangular dots, 50 nm thick, have a lateral size of 500 nm. The lattice parameter of the array is 800 nm. To study and control the magnetic behaviour of the sample we have used a commercial AFM/MFM system from Nanotec Electrónica S.L which has been conveniently modified so that magnetic fields can be applied in the course of MFM operation. Our MFM includes a PLL (Phase Lock Loop) system that keeps constant the phase of the cantilever oscillation and where the magnetic signal corresponds to the frequency shift of the cantilever oscillation. Out-of-plane and in-plane magnetic fields up to 1.5 kOe and 2 kOe, respectively, can be applied preserving the necessary high mechanical stability of the microscope[28].

The probes are commercial Si cantilevers (Nanosensors PPP –FMR, k=1.5 N/m and f = 75 KHz) coated by a CoCr sputtered thin film. Before each experiment the probes are magnetized along their pyramid axis. The thickness of the coating has been carefully selected (25 nm) in order to prevent the influence of the stray field of the tip on the magnetic state of the sample. The MFM contrast obtained with those homemade probes is relatively low (5Hz, which corresponds to a force gradient of about $10^{-4}$ N/m) but enough to identify incontestably (except in few cases) the chirality and polarity of the vortex. The behaviour of this sort of tips under an externally applied magnetic field has been thoroughly analysed recently[29].

The sample was initially demagnetized i.e., the array of nanotriangles presents the four possible vortex state configurations randomly distributed as observed in Fig 1 (b). In this MFM image, we can distinguish the MFM contrast corresponding to three in-plane domains with clockwise or counter clockwise rotation (as indicated by the arrows) and the core of the vortex with the magnetization pointing in up or down direction (black or white contrast respectively). In previous work [see ref 26], we have confirmed the lack of interactions between neighbouring triangles in this array.

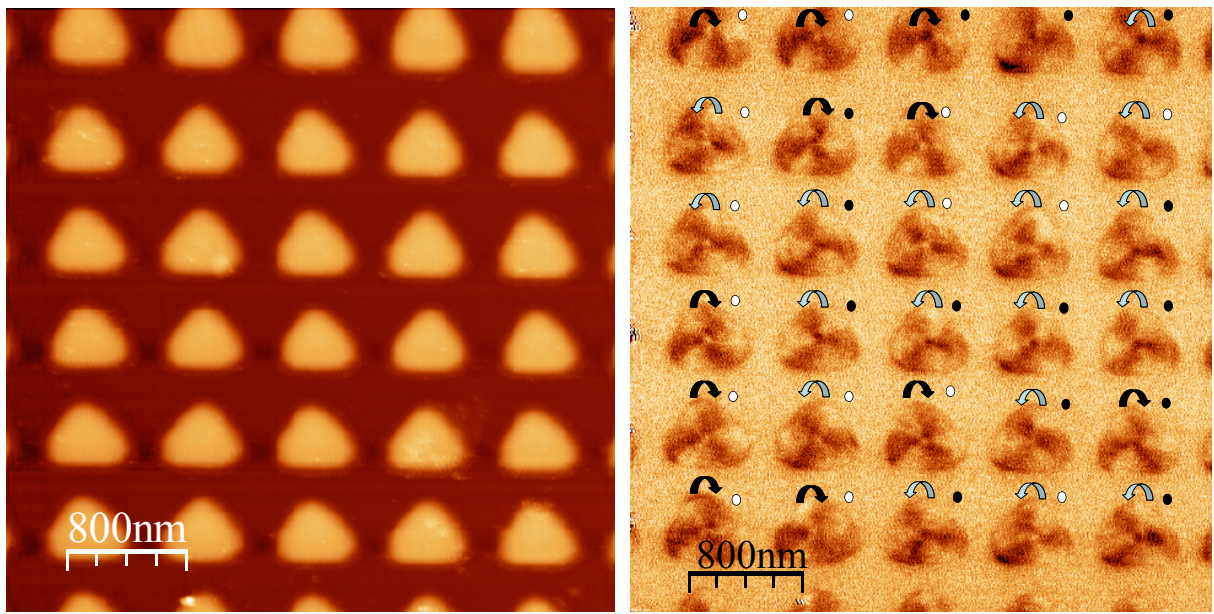

**Figure 1.** (Color online) (a) AFM and (b) MFM images of a Ni nanostructure arrangement after demagnetizing the sample. Black and white spots in the triangles of Fig 1b denote the core polarity pointing in up or down direction, respectively. The arrows represent the chirality of the closure flux.

In Figure 2 series, *in situ* magnetic field is applied along in-plane direction to induce a well defined chirality in each structure. At the initial state (Fig 2a), the four triangles present the same vortex polarity and random chiralities. When the magnetic field is applied in +x direction the vortex core moves towards the base or the top of the triangle regarding their clockwise or counter clockwise chirality respectively[26] (Fig 2b). An increase of the field large enough to nearly saturate the magnetic structures results in a single dipolar contrast (Fig 2c). Notice that upon decreasing the value of the applied field, the vortex nucleates always in the base of the triangles (Fig 2d). However, they present random vortex polarities since no out-of-plane field is applied. Finally, at zero field the resulting chirality of each nanotriangle is determined by the direction of the previously saturated magnetic state (see Fig 2f). In Figure 3 we present an image corresponding to a larger region than that of Fig 2f indicating that the vortex core chirality has been changed everywhere. The observed magnetisation-remagnetization process and the corresponding MFM images are in agreement with static micromagnetic simulations, as was reported in Ref. [26].

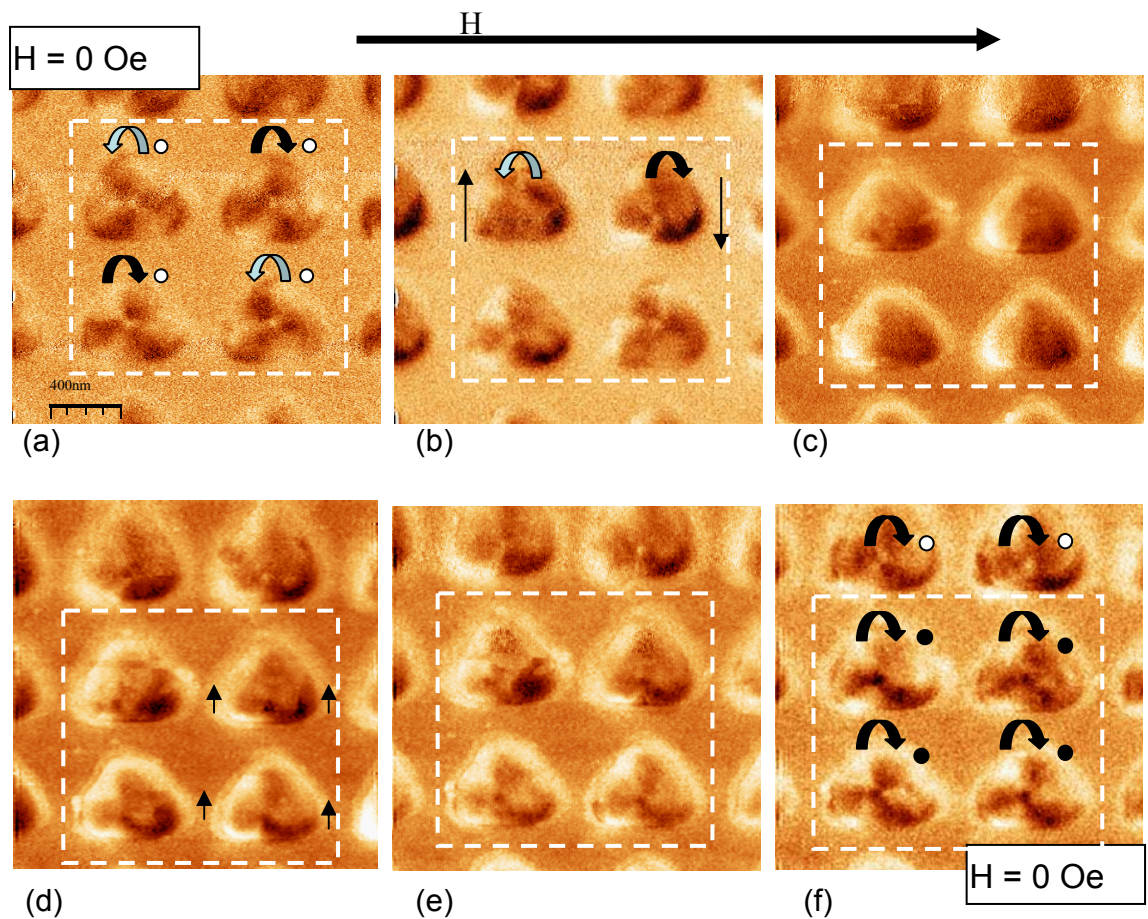

**Figure 2.** (Color online) Series of MFM images obtained at (a) 0 Oe, (b) +150 Oe, (c) +200 Oe, (d) + 100 Oe, (e) +50 Oe, (f) 0 Oe. The round arrows in the Figure schematically show the vortex chirality direction while the straight

ones – the direction of the core movement. Straight large arrow outside the MFM images indicates the direction of the applied magnetic field.

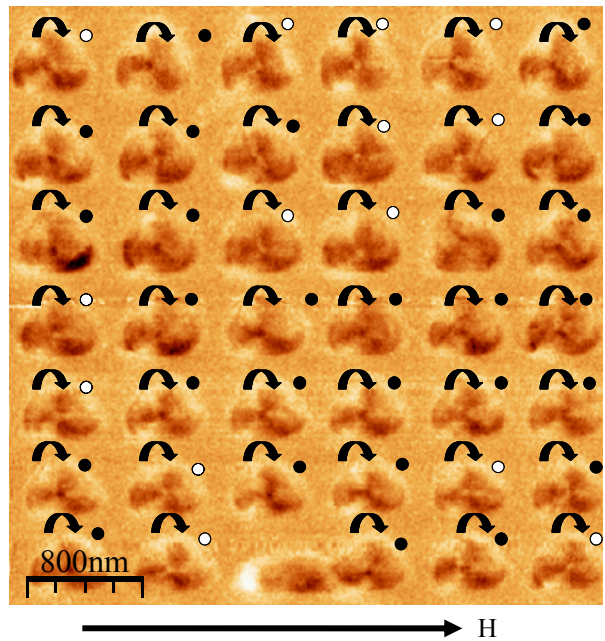

**Figure 3.** (Color online) Image obtained after saturating the sample by applying an in-plane magnetic field in +x direction (see arrow). This image is subsequent to the series in Figure 2. This image is a zoom-out of Fig 2f.

To establish the reproducibility of the method, the magnetic field is now applied in the opposite direction to produce the counter clockwise chirality of the nanostructures. Fig 4a corresponds to the MFM image obtained after saturating the sample under a magnetic field applied along the -x direction. Note that while the chirality of the closure flux is opposite to that observed after saturating along the +x direction, the vortex core is again randomly distributed in up and down directions.

As a result of our observation, the vortex chirality can be controlled by an in-plane field producing arbitrary vortex polarity.

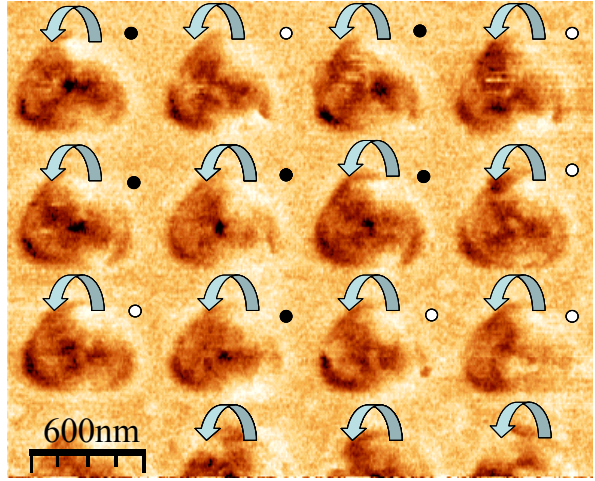

**Figure 4.** (Color online) MFM image obtained after saturating the sample in -x direction to control the counter clockwise chirality of the closure flux.

### B. Micromagnetic Simulations

To understand the micromagnetic mechanism and the minimum requirement for the control of vortex chirality and polarity we perform micromagnetic simulations. For the simulations, Ni triangles of the same dimensions as the experimental ones were discretized in finite elements. The finite element Magpar code[30] has been used for the micromagnetic simulations considering the following Ni parameters: the anisotropy constant K= -5 x $10^3$ (J/m$^3$), the exchange stiffness A= 3.4 x $10^{-12}$ (J/m), the saturation magnetisation value $\mu_0 M_s$= 0.61 (T).

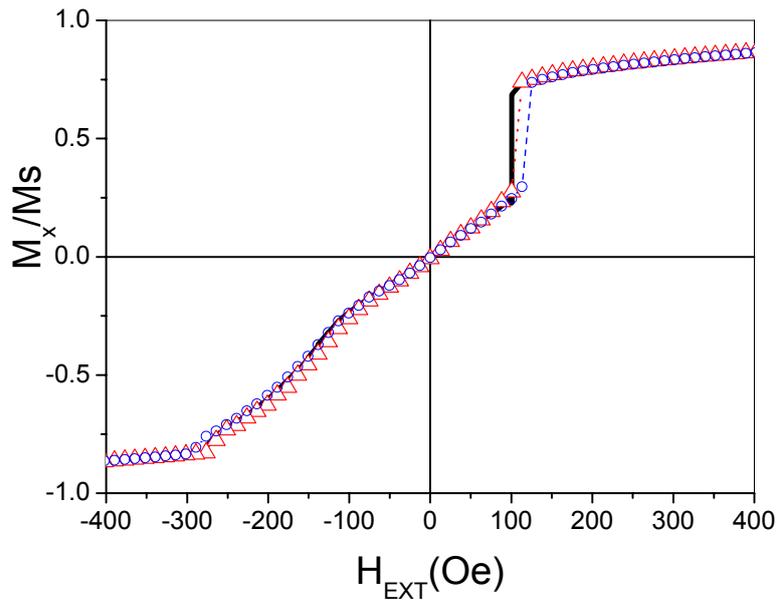

**Figure 5.** (Color online) Descending branch of the simulated hysteresis cycle for one isolated dot (solid line) and four nanodots separated by 300 nm (○) and 100 nm (Δ). The in-plane magnetic field $H_{EXT}$ is parallel to the base of the triangles.

Fig. 5 presents modelled hysteresis cycles for a single magnetic nanodot (▬) and 4 nanodots separated by 300 nm (○) (as in the experimental situation) and 100nm (Δ). The in-plane magnetic field $H_{EXT}$ is parallel to the base of the triangles. As we clearly see, the interactions in our system are negligible (the differences may be attributed to different nanodot discretizations). We conclude that for this separation the stray fields from the triangular corners have little effect on the vortex nucleation field. Consequently from now on we will focus our modelling on single dots only.

The use of variable-field MFM has enabled the experimental demonstration that chirality can be controlled in the static regime, that is, saturating the sample in a given direction by applying the field for sufficiently long time. To present a complete picture of the vortex states control, we should also consider the case of fast pulse fields which is also more interesting from the application point of view. In what follows, we will investigate the vortex configuration as a function of the applied field duration, understanding that the long pulse duration corresponds to that of the static control. For this purpose, the external in-plane field of the maximum value, $H^{max}$ is supposed to have the risetime 0.5 ns, duration, $t_H$, and the decay time 0.5ns.

As observed experimentally, long field pulses durations produce reversal of the core chirality. The mechanism is illustrated in Fig 6 and is the same in the dynamical and in the static case. The control of the core chirality is provided by the triangular geometry of the dot. Due to the energy minimization, the vortex is always created on the triangular base (as observed also by variable-field MFM). During the first part of the process the core polarisation is reversed due to creation of the vortex-antivortex pair. The new created vortex with the opposite polarisation is expelled through the triangle corner opposite to the triangle base. Before this happens, the magnetisation in the triangle starts to divide into two domains- precursor of the new chirality. It is known that the vortex core moves perpendicular to the field and the direction of the motion depends on the vortex chirality [9]. Thus, there is a coupling between the direction of the chirality and the direction of motion which in our case is always from the triangular base to the opposite tip. Therefore, unlike circular and squared dots, the chirality in triangular nanodots is completely determined by the applied field direction. For both applied field directions the vortex is created on the triangular base and moves towards its centre, where it is stabilized at zero fields.

Thus, the mechanism of the vortex chirality switching is the expulsion of the vortex core from the dot and the creation of the new vortex with the chirality compatible with the applied field direction. In other words, application of the field in +x direction selects clockwise chirality and visa versus as shown in the above experimental results (Fig 2, Fig3 and Fig4). Unlike the static case of MFM imaging, in the particular case of Fig 6, the field is switched off before the original vortex is expelled.

The process of the chirality change occurs in timescale of the order of 5 ns. For stronger and longer field pulses multiple vortex-antivortex pairs creation can be observed and the final core polarisation becomes random, similar to static MFM images.

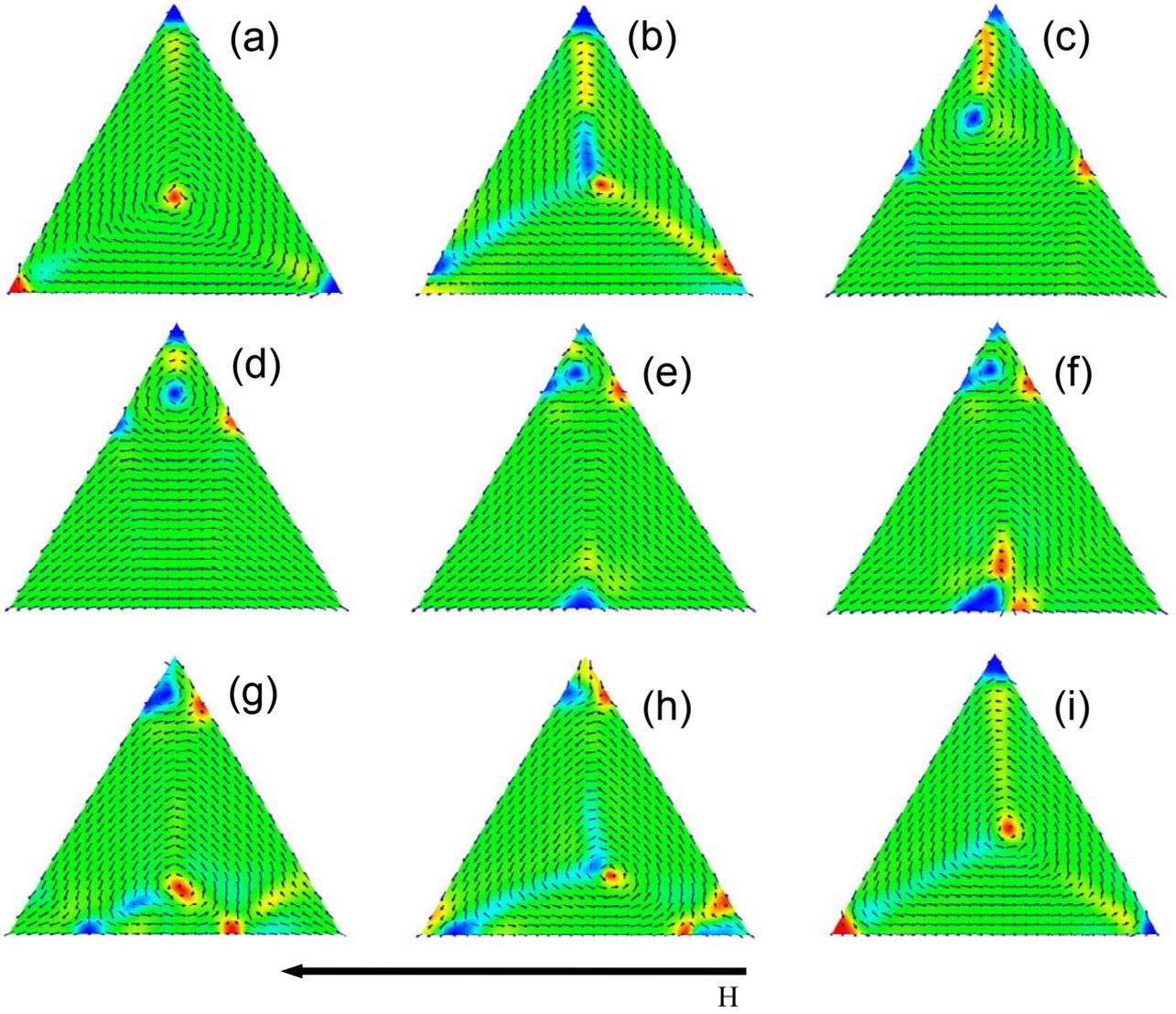

**Figure 6.** (Color online) Dynamical configurations during the vortex chirality reversal under applied field $H_{max}$=250 Oe, duration $t_H$=3 ns and field rise and decay times 0.5ns at subsequent timeshots. (a) t=0, the original vortex with core up and clockwise chirality is in the nanodot centre. (b) t=0.56ns, the vortex-antivortex pair is created. (c) t= 1.675 ns, the original vortex has annihilated with the antivortex. The new core-down vortex is moving perpendicular to the field direction towards the triangle corner. (d) t=2.68ns and (e) t= 3.9ns, the vortex continues to move towards the triangle corner. The 90-degree domain wall (an onset of the conter clockwise chirality) is created (f) t=4.1ns. The domain wall gives birth to two vortices with the opposite polarisations and an anti-vortex (g) t=4.5ns The vortex (core down)-antivortex pair is annihilated. (h) t=4.83 ns, the vortex moves perpendicular to the field towards the nanotriangle centre. (i) t=5.63 ns, the vortex is stabilised in the nanodot centre.

We have observed that for short duration field pulses only vortex core polarity is changed. The mechanism of core polarisation switching is the same to that reported earlier for circular dots[13,14,15]. This mechanism is illustrated in Fig 7. Namely, a new vortex-antivortex pair is created with the subsequent annihilation of the original vortex and anti-vortex (see Fig 7b). The remaining new vortex with changed polarisation performs a circular motion around the dot centre (Fig 7 c-e) until this motion is damped out (Fig 7f). Note that the process of the core polarisation reversal occurs in the timescale less than 0.5ns, after this moment the applied field may be switched off.

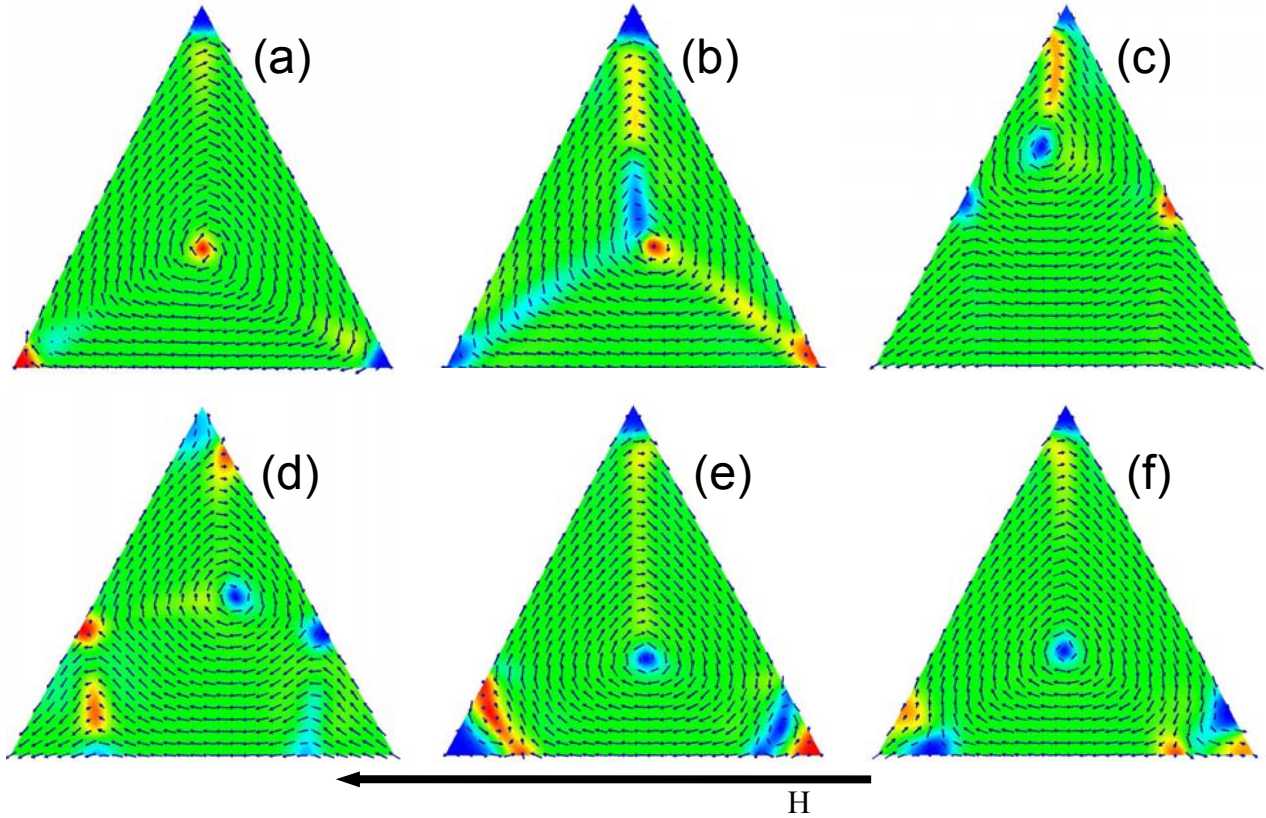

**Figure 7.** (Color online) Dynamical configurations during the vortex polarisation reversal under applied field $H_{max}$=250 Oe and duration $t_H$=1ns at subsequent timeshots (a) t=0 ns Original vortex with the polarisation up is in the nanodot centre. (b) t=0.5ns Creation of the vortex-antivortex pair (c) t=1.5 ns The original vortex has disappeared, the new vortex with the polarisation down starts to move around the nanodot centre in the clock-wise direction. (d) t=2.7ns and (e) t=3.63ns Continuation of the vortex gyration (f) t=4.64 ns. The gyroscopic motion is damped out and the vortex is stabilised in the dot centre.

In Fig 8 we have presented the diagram showing the results of the application of external fields applied along –x direction of different durations and strengths. Notice that in some region of the parameters the polarisation of the new vortex is determined by the polarisation of the previous one and is opposite to it due to the magnetostatic energy minimization. However, the magnetostatic coupling coming from the vortices cores is negligible and the field parameters range where we observed that both the polarity and the chirality could be controlled is small.

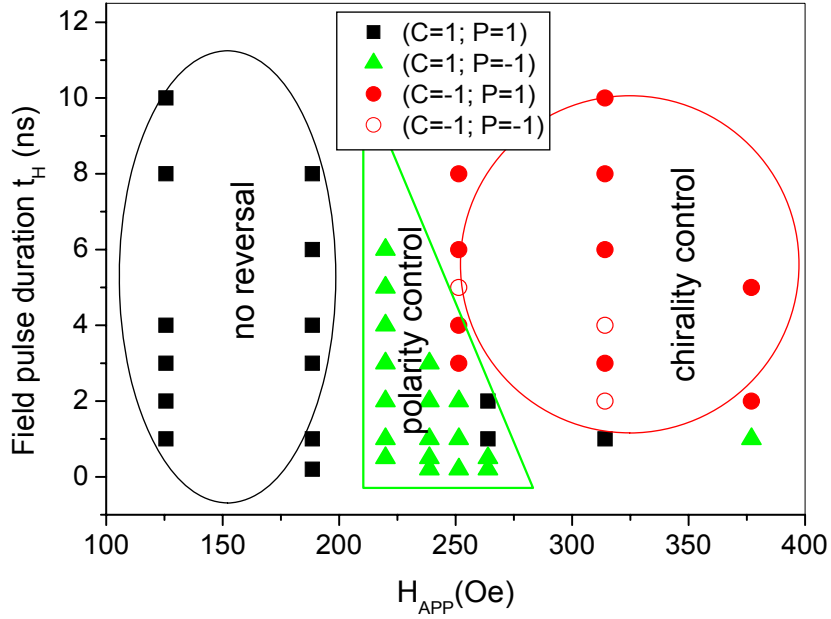

**Figure 8.** (Color online) Diagram showing the result of the application of the external field with the maximum strength $H^{max}$, field pulse duration $t_H$ and field rise and decay times 0.5ns. The initial vortex state is core-up polarisation (P=1) and clockwise chirality (C=1). The final states are indicated by different symbols corresponding to clockwise (C=1) or counter clockwise (C=-1) chiralities and polarisation up (P=1) or down (P=-1).

To explore new possibilities of chirality and polarity vortex control we propose to combine in-plane and out-of-plane magnetic fields.

III. CHIRALITY AND POLARITY CONTROL BY A COMBINED FIELD

**A. Experimental results**

The polarity of the vortex can be selected by applying an external out-of-plane field as shown in Fig 9a. Vortex is quite a stable object and demagnetising field from the thin film geometry is quite strong, consequently, strong perpendicular fields are necessary (around 700 Oe) to change the polarity of the vortex. However, such perpendicular fields will leave chirality undefined as shown in the MFM image in Fig 9a. It could be expected that with a combined action of out-of-plane and a variable in-plane fields, one can independently control vortex chirality and polarity. Experimental evidences have been found of such vortex configuration control. MFM image in Fig 9b has been acquired in remanent state after applying a magnetic field of 1500 Oe at 30º from the surface plane. Both chirality and polarity were determined for all individual triangles through the combined action of static in-plane and out-of-plane fields applied out of the MFM system. It is reasonable to suppose that since the vortex appears on the triangular base with arbitrary polarity, a small component of the perpendicular field could help to control vortex polarity. However, experimentally it has been found that the required field is not small. For example, the magnetic field produced by the MFM tip (about 200 Oe) is not sufficient to stabilize the nucleated vortex core with a desired direction as deduced from the results shown in Fig3 and Fig 4.

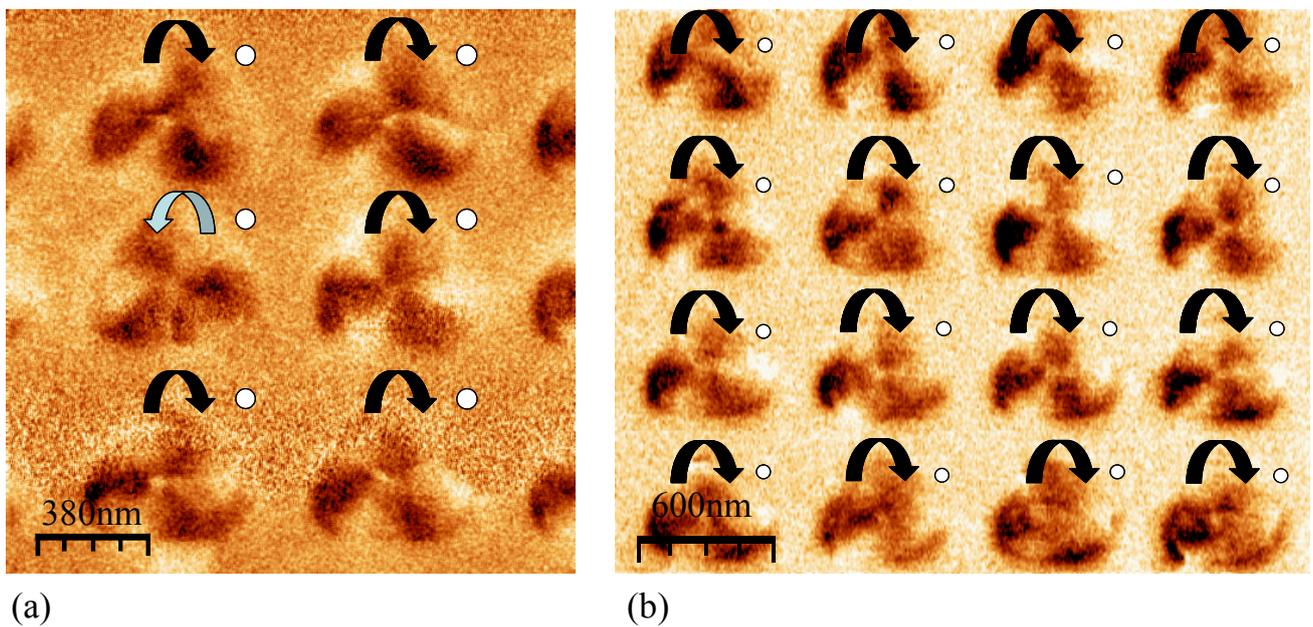

(a) (b)

**Figure 9.** (Color online) (a) MFM image obtained after saturating the sample in out-of-plane direction to control the vortex polarity. (b) MFM image obtained after applying a combined in-plane and out-of-plane field producing the clockwise configured nanostructures with vortex cores pointing down.

### B. Micromagnetic Simulations

The experimental results suggest that one can help the creation of the vortex with specific polarisation applying a magnetic field with an out-of-plane component. To investigate this idea, we have performed micromagnetic simulations of hysteresis cycles with applied fields at different angles for the case of four magnetic nanodots. The micromagnetic simulations clarify us the vortex behaviour.

Fig.10 represents simulated hysteresis cycles (corresponding to the magnetization along the applied field direction) in a system of 4 nanodots and shows that the vortex nucleation field increases with the out-of-plane field angles. Since we use finite element method, each of our triangles is different, emulating the experimental situation. The results show that in the remanence corresponding to Fig.11 (in-plane field), three of our triangles appeared with vortex core configurations "up" and one "down". Saturating our sample with field applied at 5º out-of-plane and having a negative component we have obtained at the remanence two vortices with core "up" and two with core "down". It was necessary to apply a field with at least 35º out-of-plane to obtain all the vortices in the configuration core "down".

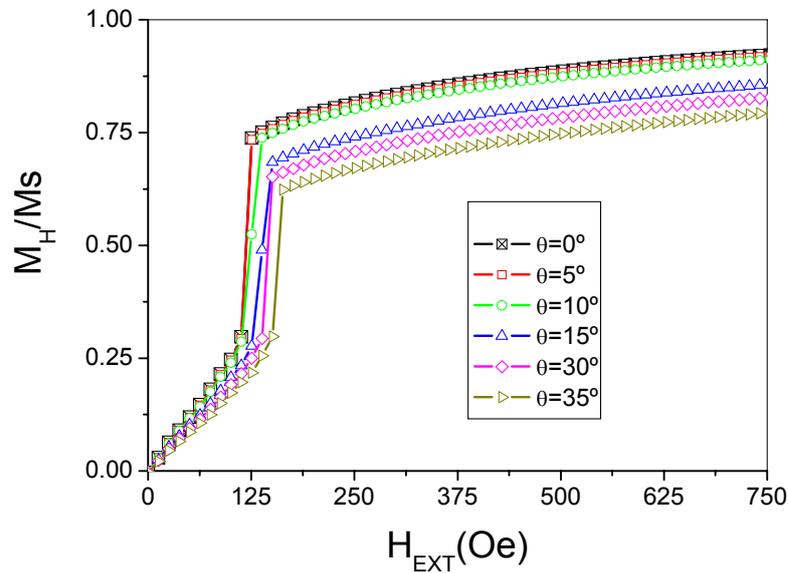

**Figure 10.** (Color online) Part of the descending branch of the simulated hysteresis cycles with fields applied at different out-of-plane angles, θ, for a system of four dots separated by 300 nm.

To understand the situation, we present in Fig.11 the simulated MFM images (a, c, d and f) and the color map plot of the $M_z$ component (b and e) before and after the vortex nucleation for applied fields at 5º (a-c) and 35º (d-f). The simulated MFM images (Fig.11 a and d) show that before the vortex polarity formation the chirality is already present, as was also mentioned above. Because of this, the local field on the triangular base is already parallel to it and, therefore, the action of the perpendicular external field is reduced by the dipolar field. Consequently, the overall internal field is not sufficient to produce the desired polarisation. Additionally it is worth to mention that in the upper left triangle the original vortex was created with the polarisation down but dynamically changed it, as it is seen in Fig.11 (c).

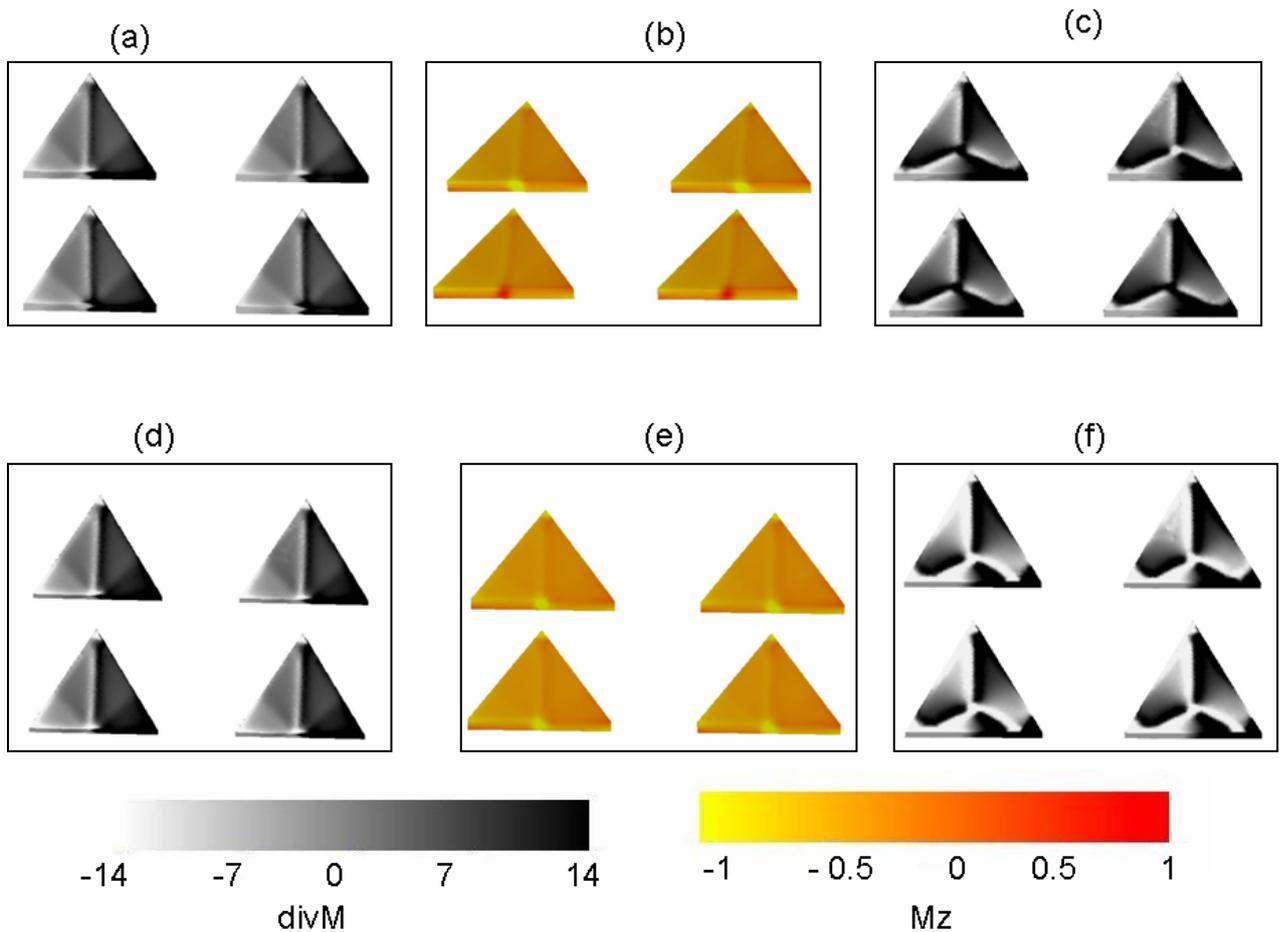

**Figure 11.** (Color online) Simulated images of the four triangular nanodots with applied field at 5º (upper rows, a-c) and 35º (lower rows, d-f). The left and middle columns are images before the vortex nucleation, (a and d) represent the simulated MFM images and (b and e) are the corresponding Mz component. The right column presents the simulated MFM images after the vortex nucleation.

For the field applied at 35º the nucleation field is larger than that for 5º, so that it's perpendicular component and the overall balance of the total field at this moment is negative, allowing the creation of the vortex polarisation in the desired direction. To summarise, the simulations show that in order to achieve the vortex polarisation control, at the moment and in the place of the vortex creation the overall total field should be pointed in the desired direction and this could be achieved only with out-of-plane field components of about 430 Oe.

IV. CONCLUSIONS

In conclusion, by means of variable field MFM and micromagnetic simulations we have shown that triangular nanodots offer a possibility to simultaneously control magnetic vortex chirality and polarity. The four magnetic states are perfectly visible by Magnetic Force Microscope offering the possibility for codification.

We have studied theoretical and experimentally the effect of in-plane magnetic fields applied parallel to the triangle base. In the static case, we observe the change of the vortex chirality due to the vortex annihilation and the coupling of the vortex core direction motion and the applied field direction.

We have shown that in a dynamic regime, by varying the field pulse duration it is possible to control both vortex chirality and polarity: with short field pulses it is possible to change the vortex polarity while with longer pulses the chirality can be changed. The mechanism of the vortex chirality control in this conditions is the same as that in the static case. At the same time, the mechanism for the change of the vortex polarisation for short field pulses is due to the vortex-antivortex pair creation. Thus, by triggering pulses (in-plane direction) with different durations one can control the four vortex states.

Alternatively, the change of the vortex polarisation can be produced by applying additionally a perpendicular field component. Since the vortex chirality is formed before its polarity, the perpendicular applied field action on the triangle base is reduced by the magnetostatic field. The micromagnetic simulations show that for a complete polarity control a minimum strength of the out-of-plane component (approx. 430 Oe) is necessary. The corresponding experimental value could be larger due to the magnetostatic shape anisotropy arising from the thin film geometry.

For the future applications, the use of dynamically induced polarity control has some advantages as compared to that of the static fields. The control of the vortex states makes the triangular dots unique candidates for the applications in non-volatile magnetic storage such as vortex MRAM memories.

ACKNOWLEDGMENTS: The authors acknowledge the financial support from the Spanish Ministerio Ciencia e Innovación through the projects NAN2004-09087, NAN2004-09183-C10-04, FIS2005-07392, FIS2008-06249, MAT2007-66719-C03-01, MAT2007-65420-C02-01, Consolider CSD2007-00010, CS2008-023 and CAM grants S-0505/ESP/0337, 505/MAT/0194 and CAM-CSIC200660M046, and Fondo Social Europeo. M.Jaafar and R.Yanes wish to thank CAM and CSIC, respectively for the financial support.

REFERENCES:

[1] J.I. Martin, J. Nogues, K. Liu, J.L. Vicent, I.K. Schuller, J.Magn.Magn.Mater. **256**, 449 (2003)

[2] B.Azzerboni, L.Pareti, G.Asti "Magnetic Nanostructures in Modern Technology: Spintronics, Magnetic MEMS and Recording", Soringer-Verlag, New York, (2007).

[3] R.P. Cowburn, J. Phys.D: Apply.Phys. **33**, R1-R16 (2000)

[4] J.L. Costa-Kramer, J.I. Martin, J.L. Menendez, A.Cebollada, J.V.Anguita, F. Briones and J.L. Vicent, Appl. Phys. Lett. **76**, 3091 (2000).

[5] H.Hoffman and F.Steinbauer, J. Appl. Phys. **92**, 5463 (2002)

[6] K.L. Metlov and K.Y. Guslienko, J.Magn.Magn.Mater. **242-245**, 1015 (2002)

[7] H. Stoll, A. Puzic, B. van Waeyenberge, P. Fischer, J. Raabe, M. Buess, T. Haug, R. Höllinger, C. Back, D. Weiss, and G. Denbeaux, Appl. Phys. Lett. **84**, 3328–3330 (2004).

[8] B.Van Wayenberge, A.Puzic, H.Stoll, K.W.Chou, T.Tyliszczak, R.Hertel, M.Farle, H.Brukl, K.Rott, G.Reiss, L.Neudecker, D.Weiss, C.H.Back and G.Shutz, Nature (London) **444**, 461 (2006).

[9] W.Scoltz, K.Yu.Guslienko, V.Novosad, D.Suess, T.Schrefl, R.W.Chantrell and J.Fidler, J.Magn.Magn.Mater. **266**, 155 (2003).

[10] A.Thiaville, J.M.Garcia, R.Dittrich, J.Miltat and T.Schrefl, Phys. Rev. B **67**, 094410 (2003).

[11] K.Yamada, S.Kasai, Y.Nakatani, K.Kobayashi, H.Kohno, A.Thiaville and T.Ono, Nature Mater. **6**, 270 - 273 (2007).

[12] R.P.Cowburn, Nature Mater. **6**, 255 (2007).

[13] K.S.Lee, K.Y.Guslienko, J.Y.Lee and S.-K.Kim, Phys Rev. B **76**, 174410 (2007).

[14] R.Hertel, S.Gliga, M.Fahnle and C.M.Schenider, Phys Rev Lett **98**, 117201 (2007).

[15] K.Y.Guslienko, J.-Y.Lee and S.K.Kim, Phys Rev Lett **100**, 027203 (2008).

[16] J.Thomas, Nature Nanotechn., **2**, 206 (2007).

[17] S.-K.Kim, K.-S.Lee, Y.-S.Yu, Y.-S.Choi, Appl. Phys. Lett. **92**, 022509 (2008).

[18] M. Schneider, H. Hoffmann, and J. Zweck, Appl. Phys. Lett. **79**, 3113 (2001)

[19] Kuo-Ming Wu, Jia-Feng Wang, Yin-Hao Wu, Ching-Ming Lee, Jong-Ching Wu, and Lance Horng, J. Appl. Phys. **103**, 07F314 (2008)

[20] T. Taniuchi, M. Oshima, H. Akinaga, K. Ono, J. Appl. Phys. **97**, 10J904 (2005)

[21] P. Vavassori, R. Bovolenta, V. Metlushko, B. Ilic, J. Appl. Phys. **99**, 053902 (2006)

[22] F. Giesen, J. Podbielski, B. Botters and D. Grundler, Phys. Rev. B **75**, 184428 (2007)

[23] P. Vavassori, O. Donzelli, M. Grimsditch, V. Metlushko, and B. Ilic, J. Appl. Phys. **101**, 023902 (2007)

[24] P. Vavassori, D. Bisero, V. Bonanni, A. Busato, M. Grimsditch, K. M. Lebecki, V. Metlushko, and B. Ilic, Phys. Rev. B **78**, 174403 (2008)

[25] S. Y. H. Lua, S. S. Kushvaha, Y. H. Wu, K. L. Teo, and T. C. Chong, Appl. Phys. Lett. **93**, 12250 (2008)

[26] M. Jaafar R. Yanes, A. Asenjo, O. Chubykalo-Fesenko, M. Vázquez, E. M. González and J. L. Vicent, Nanotechnology **19**, 285717 (2008).

[27] J.I. Martin, Y. Jaccard Y, A. Hoffmann, J.Nogues, J.M. George, J.L.Vicent and I.K. Schuller, J. Appl. Phys. **84**, 411 (1998).

[28] M. Jaafar, J Gómez-Herrero, A Gil, P Ares, M Vázquez and A Asenjo, Ultramicroscopy, **109**, 693 (2009)

[29] M. Jaafar, A. Asenjo and M. Vázquez, IEEE Nano **7**, 245(2008)

[30] http://magnet.atp.tuwien.ac.at./scholz/magpar